\documentclass{article}%
\usepackage{amsfonts}
\usepackage{amsmath}
\usepackage{amssymb}
\usepackage{graphicx}%
\begin{document}

\title{Casimir Effect in the Presence of Minimal Lengths }
\author{Kh. Nouicer$\thanks{E-mail: khnouicer@mail.univ-jijel.dz} $\\Laboratory of Theoretical Physics and Department of Physics,\\Faculty of Sciences, University of Jijel,\\Bp98, Ouled Aissa, 18000 Jijel, Algeria.}
\date{}
\maketitle

\begin{abstract}
It is expected that the implementation of minimal length in quantum models
leads to a consequent lowering of Planck's scale. In this paper, using the
quantum model with minimal length of Kempf et al \cite{kempf0}, we examine the
effect of the minimal lenght on the Casimir force between parallel plates.

\end{abstract}


\section{Introduction}

The construction of a quantized theory which incorporate gravity remains one
of the priorities of theoretical physicists. Unfortunately all the attempts
toward this goal fail. The reason is that the Planck scale $l_{\text{p}%
}=1.61605\times10^{-35}%
\operatorname{m}%
$, at which the effects of quantum gravity reveal themselves is so small, that
these effects are neglected in experimentally accessible energies. Recently,
to cure this problem, different scenarios have been proposed and all leading
to a significant lowering of Planck's scale. Among them, models with large
eXtra dimensions (LXD) \cite{arkani}, non commutative field theory models
\cite{nekrasov} and models with non zero minimal lengths \cite{kempf0}. In
this paper we are interested in the later models based on generalized
commutation relations $\left[  \hat{x}_{i},\hat{p}_{j}\right]  =i\hbar\left[
\left(  1+\beta\hat{p}^{2}\right)  \delta_{ij}+\beta^{\prime}\hat{p}_{i}%
\hat{p}_{j}\right]  .$ These commutations relations lead to a generalized
uncertainty principle (GUP) which define non zero minimal lengths in position
and/or momentum. A non zero minimal length in position has first appeared in
the context of perturbative string theory \cite{groos}. One major feature of
this finding is that the physics below such a scale becomes inaccessible and
then define a natural cut-off which prevents from the usual UV divergencies.
The other consequence of such GUP is the appearance of an intriguing UV/IR
mixing, first noticed in the ADS/CFT correspondence \cite{suskind}. Physically
the UV/IR mixing means that we can probe short distances physics by long
distances physics. We point that the UV/IR mixing is also a feature of non
commutative quantum field theory \cite{nekrasov,micu}. On the other\ hand some
scenarios have been proposed where non zero minimal length is related to large
eXtra dimensions \cite{hos01}, to the running coupling constant \cite{hos02}
and to the physics of black holes production \cite{hos03}.

Recently the cosmological constant problem and the classical limit of the
physics with minimal length have been investigated by the group of Virginia
Tech \cite{chang,chang01}. In \cite{chang01} the value of the minimal length
is so small that it seems meaningful. The size of the minimal length have been
also extracted from the energy spectrum of the Coulomb potential
\cite{brau,akhoury} and from the energy spectrum of electrons in a trap
\cite{minic}. \ 

On the other hand the Casimir force has been calculated in a model
incorporating one large eXtra dimension \cite{hossen}. The comparison with
available experimental data gives $R\lesssim10nm$ where $R$ is the size of the
compactified eXtra dimension. Motivated by the fact  that large eXtra
dimensions and minimal lengths models aim to lower Planck's scale and can be
related to each others we calculate in this paper, the effect of the presence
of a minimal length on the Casimir force between parallel plates.

The rest of the paper is organized as follows. In section II, implementing the
minimal length using standard methods of quantum mechanics we obtain
generalized uncertainty principle (GUP), generalized plane waves and modified
closure relations. In section III, we quantify the electromagnetic field and
then following the standard recipe we calculate the Casimir force between to
parallel plates. Section IV is left for concluding remarks. \ 

\section{Quantum mechanics with generalized Heisenberg relation}

Following \cite{kempf0} we consider the following realization of the position
and momentum operators \ %

\begin{equation}
X_{i}=i\hbar\lbrack(1+\beta\mathbf{p}^{2})\frac{\partial}{\partial p_{i}%
}],\qquad P_{i}=p_{i},\quad.
\end{equation}
where $\beta$ is a small positive parameter. This representation leads to the
following generalized commutators
\begin{align}
\left[  X_{i},P_{j}\right]   &  =i\hbar\delta_{ij}\left(  1+\beta
\mathbf{p}^{2}\right)  ,\\
\left[  X_{i},X_{j}\right]   &  =2i\hbar\beta\left(  P_{i}X_{j}-P_{j}%
X_{i}\right)  ,\\
\left[  P_{i},P_{j}\right]   &  =0.
\end{align}
and the generalized uncertainty principle (GUP)%

\begin{equation}
\left(  \Delta X_{i}\right)  \left(  \Delta P_{i}\right)  \geq\frac{\hbar}%
{2}\left[  1+\beta(\Delta\mathbf{p})^{2}\right]  . \label{heis}%
\end{equation}
The peculiarity of relation $\left(  \ref{heis}\right)  $ is that it exhibits the
UV/IR mixing phenomenon which allows to probe short distance physics (UV) from
long distance one (IR). A minimization of $\left(  \ref{heis}\right)  $ with
respect to $\left(  \Delta P_{i}\right)  $ gives the following non zero
minimal length%
\begin{equation}
\left(  \Delta X_{i}\right)  _{min}=\hbar\sqrt{\beta}. \label{min}%
\end{equation}
Eq.$\left(  \ref{min}\right)  $, like the UV/IR mixing, reveals the non local
character of models based on Eqs.(1-3). Then we have not localized
eigenfunctions in the $\mathbf{r}$-space. So, any eigenvalue problem can be
solved by going to the momentum space.\newline\qquad In the following we
derive necessary relations for our calculation taking in mind that we must
recover the usual quantum mechanics in the limit $\beta\rightarrow0$. First we
assume that $\mathbf{R}\mid\mathbf{r}>=\mathbf{r}\mid\mathbf{r}>$ where the
vectors $\mid\mathbf{r}>$ represent maximally localized states. They are
normalized states unlike the ones of ordinary quantum mechanics.

Using these maximally localized states we derive \ the following
quasi-position eigenvectors
\begin{equation}
f_{\mathbf{p}}\left(  \mathbf{r}\right)  =\frac{1}{\sqrt[3]{2\pi\hbar}}%
\exp\left(  -\frac{i\mathbf{r}}{\hbar\sqrt{\beta}}\arctan\mathbf{p}\sqrt
{\beta}\right)  \quad. \label{plane}%
\end{equation}
with the following generalized dispersion relation%

\begin{equation}
\lambda\left(  \left\vert \mathbf{p}\right\vert \right)  =\frac{2\pi\hbar
\sqrt{\beta}}{\arctan\left(  \mid\mathbf{p\mid}\sqrt{\beta}\right)  }\text{.}
\label{dispersion}%
\end{equation}
The states given by $\left(  \ref{plane}\right)  $ are far from being the well
known plane waves. However in the limit $\beta\rightarrow0$ we recover the
usual planes waves of ordinary quantum mechanics.\newline Now assuming the
usual closure relation for the maximally localized eigenstates $1=\int
_{-\infty}^{+\infty}d\mathbf{r}^{\prime}\mid\mathbf{r}><\mathbf{r}\mid$, we
obtain
\begin{equation}
<\mathbf{p}^{\prime}\mid\mathbf{p}>=\sqrt{\beta}\delta\left(  \arctan
\sqrt{\beta}\mathbf{p}-\arctan\sqrt{\beta}\mathbf{p}^{\prime}\right)  .
\end{equation}
Using the relation $\delta f(x)=\Sigma_{i}\frac{\delta(x-x_{i})}{f^{\prime
}(x_{i})}$, where $x_{i}$ are the roots of $f(x)$, we finally get%
\begin{equation}
<\mathbf{p}^{\prime}\mid\mathbf{p}>=(1+\beta\mathbf{p}^{2})^{\frac{1}{2}%
}(1+\beta\mathbf{p}^{\prime}{}^{2})^{\frac{1}{2}}\delta(\mathbf{p}%
-\mathbf{p}^{\prime}). \label{scalar}%
\end{equation}
From this equation we derive the modified completeness relation for the
momentum eigenstates $\mid\mathbf{p}>$
\begin{equation}
\int\frac{d\mathbf{p}}{(1+\beta\mathbf{p}^{2})}\mid\mathbf{p}><\mathbf{p}%
\mid=\mathbf{{1}.} \label{clos}%
\end{equation}
Here we observe a squeezing of the momentum space at high momentum. Let us end
our calculations by showing that the states $\mid\mathbf{r}>,$ like the
coherent states, do not form an orthogonal set. Indeed we have%

\begin{align}
<\mathbf{r}\mid\mathbf{r}^{\prime}>  &  =\int\frac{d\mathbf{p}}{(1+\beta
\mathbf{p}^{2})}f_{\mathbf{p}}\left(  \mathbf{r}\right)  f_{\mathbf{p}}^{\ast
}\left(  \mathbf{r}^{\prime}\right) \nonumber\\
&  =\int\frac{d\mathbf{p}}{(1+\beta\mathbf{p}^{2})}\exp\left\{  -\frac
{i(\mathbf{r-r}^{\prime})}{\hbar\sqrt{\beta}}\arctan\sqrt{\beta}%
\mathbf{p}\right\} \label{coh}\\
&  =\frac{1}{\pi(\mathbf{r-r}^{\prime})}\sin\left(  \frac{\pi(\mathbf{r-r}%
^{\prime})}{2\hbar\sqrt{\beta}}\right)  .\nonumber
\end{align}
The right hand is a well behaved function unlike the Dirac distribution of
ordinary quantum mechanics. It is clear that the limit $\beta\longrightarrow0$
restores the usual normalization $<\mathbf{r}\mid\mathbf{r}^{\prime}%
>=\delta(\mathbf{r-r}^{\prime})$. In conclusion we have chosen to work with
the normalization constant $1/\sqrt[3]{2\pi\hbar}$, while this choice renders
the states given by Eq. $\left(  \ref{plane}\right)  $ unphysicals, to
reproduce in the limit $\beta\rightarrow0$ the usual results of quantum
mechanics.\bigskip

\section{Casimir effect}

The most general solution of Maxwell equations in the presence of a minimal
length in the Coulomb gauge for slowly moving particles is given by%

\begin{equation}
\mathbf{\hat{A}}\left(  \mathbf{r},t\right)  =\sqrt{2\pi\hbar^{3}c\sqrt{\beta
}}\int\frac{d\mathbf{p}}{\left(  1+\beta\mathbf{p}^{2}\right)  \sqrt
{\arctan{\sqrt{\beta}\mid\mathbf{p}\mid}}} \sum_{\gamma=\pm1}\left[
f_{\gamma}\left(  \mathbf{p},\omega\right)  \hat{a}_{\gamma}\left(
\mathbf{p}\right)  +f^{\ast}\left(  \mathbf{p},\omega\right)  \hat{a}_{\gamma
}^{\dagger}\left(  \mathbf{p}\right)  \right]
\end{equation}
where $f_{\gamma}\left(  \mathbf{p},\omega\right)  $ are generalized plane
waves which can be obtained from Eq.$\left(  \ref{plane}\right)  $%

\begin{equation}
f_{\gamma}\left(  \mathbf{p},\omega\right)  =\frac{\varepsilon_{\gamma}\left(
\mathbf{p}\right)  }{\sqrt[3]{2\pi\hbar}}\exp\left(  \frac{i}{\hbar\sqrt
{\beta}}\left[  \mathbf{r}\arctan\left(  \mathbf{p}\sqrt{\beta}\right)
-\hbar\omega\left(  \mid\mathbf{p}\mid\right)  t\right]  \right)  , \label{pp}%
\end{equation}
with $\omega\left(  \mid\mathbf{p}\mid\right)  $ defined by the generalized
dispersion relation $\left(  \ref{dispersion}\right)  $ and $\varepsilon
_{\gamma}\left(  \mathbf{k}\right)  $ are the polarization vectors verifying%

\begin{equation}
\mathbf{\varepsilon}_{\gamma}\left(  \mathbf{p}\right)  \mathbf{\varepsilon
}_{\gamma^{\prime}}^{\ast}\left(  \mathbf{p}\right)  =\delta_{\gamma
\gamma^{\prime}}.
\end{equation}
From $\left(  \ref{pp}\right)  $ we derive the following normalization condition%

\begin{equation}
\int d\mathbf{r}f_{\gamma}^{\ast}\left(  \mathbf{p},\omega\right)
i\overleftrightarrow{\partial}_{0}f_{\gamma^{\prime}}\left(  \mathbf{p}%
^{\prime},\omega^{\prime}\right)  =\delta_{\gamma\gamma^{\prime}}%
(1+\beta\mathbf{p}^{2})^{\frac{1}{2}}(1+\beta\mathbf{p}^{\prime}{}^{2}%
)^{\frac{1}{2}}\delta(\mathbf{p}-\mathbf{p}^{\prime})
\end{equation}
The creation and annihilation operators are non relativistic ones and, since
the momentum operators are commuting, \ they satisfy the usual commutation relation,%

\begin{equation}
\left[  \hat{a}_{\gamma}\left(  \mathbf{p}\right)  ,\hat{a}_{\gamma^{\prime}%
}^{\dagger}\left(  \mathbf{p}^{\prime}\right)  \right]  =\delta_{\gamma
\gamma^{\prime}}\delta(\mathbf{p}-\mathbf{p}^{\prime}).
\end{equation}
This result, along with Eq.$\left(  \ref{coh}\right)  ,$ can be used to derive
a modified commutation relation between the fields%

\begin{equation}
\left[  A^{i}\left(  \mathbf{r},t\right)  ,E^{j}(\mathbf{r}',t)\right]
=i\left(  \delta^{ij}-\frac{\partial^{i}\partial^{j}}{\nabla^{2}}\right)
\frac{\sin\left(  \frac{\pi(\mathbf{r-r}^{\prime})}{2\hbar\sqrt{\beta}%
}\right)  }{\pi(\mathbf{r-r}^{\prime})}.
\end{equation}
Using the well known relation $\delta\left(  x\right)  =\lim_{\varepsilon
\rightarrow0}\frac{\sin\left(  x/\varepsilon\right)  }{\pi x}$ we obtain the
usual commutation relation in the limit $\beta\rightarrow0$.

Armed with this background, let us then attack the Casimir effect with square
parallel plates of sides $L$. Then the electromagnetic field must satisfy
boundary conditions. In our case we have from $\left(  \ref{pp}\right)  $%

\begin{equation}
\kappa_{3}=\frac{\hbar n\pi}{a}, \label{geo}%
\end{equation}
where $a$ is the plates separation, $\kappa_{3}=\frac{p_{3}}{\sqrt{\beta\left(  \mathbf{q}^{2}+p_{3}%
^{2}\right)  }}\arctan\left(  \sqrt{\beta\left(  \mathbf{q}^{2}+p_{3}%
^{2}\right)  }\right)  $ and $\mathbf{q}$ is the transverse momentum along the
plates. In $\left(  \ref{geo}\right)  $ we have a finite number of modes
$n=0,1,2,..,n_{\text{max}}=\left[  \frac{a}{2\left(  \Delta x\right)
_{\text{min}}}\right]  $ \ where $\left[  \cdots\right]  $ denotes the next
smaller integer. Then the geometrical quantization given by $\left(
\ref{geo}\right)  $ fulfills the requirement that in quantum models with a
minimal length, Compton wavelength cannot take arbitrary values. Indeed we
have $\lambda_{\min}$ $=4\hbar\sqrt{\beta}.$

Since $\beta$ is a small parameter we have tried a series solution to the
eight order in $\beta.$ In the following we just show the following truncated solution%

\begin{equation}
p_{3}\left(  n\right)  =\frac{\hbar n\pi}{a}\left[  1+\frac{\beta}{3}\left(
\mathbf{q}^{2}+\left(  \frac{\hbar n\pi}{a}\right)  ^{2}\right)  +\frac
{\beta^{2}}{45}\left(  2\left(  \mathbf{q}\frac{\hbar n\pi}{a}\right)
^{2}-4\mathbf{q}^{4}+12\left(  \frac{\hbar n\pi}{a}\right)  ^{4}\right)
+\cdots\right]  . \label{cond}%
\end{equation}
In figure 1 we have plotted the modified wavelengths associated with momentums
$\kappa_{3}$ and $p_{3}$ to the eight order in $\beta$ for $\beta=0.01$ and
$\hbar=a=q=1$. For large $n$ the wavelength associated with $\kappa_{3}$ tends
asymptotically to $\lambda_{\min}$ while the one associated with $p_{3}$ tends
to zero faster than the wavelength of the usual theory. A similar behavior has
been obtained in \cite{kemp4} using generalized dispersion relations.

\begin{figure}
\begin{center}
\includegraphics[height=10.0364cm, width=10.0364cm]{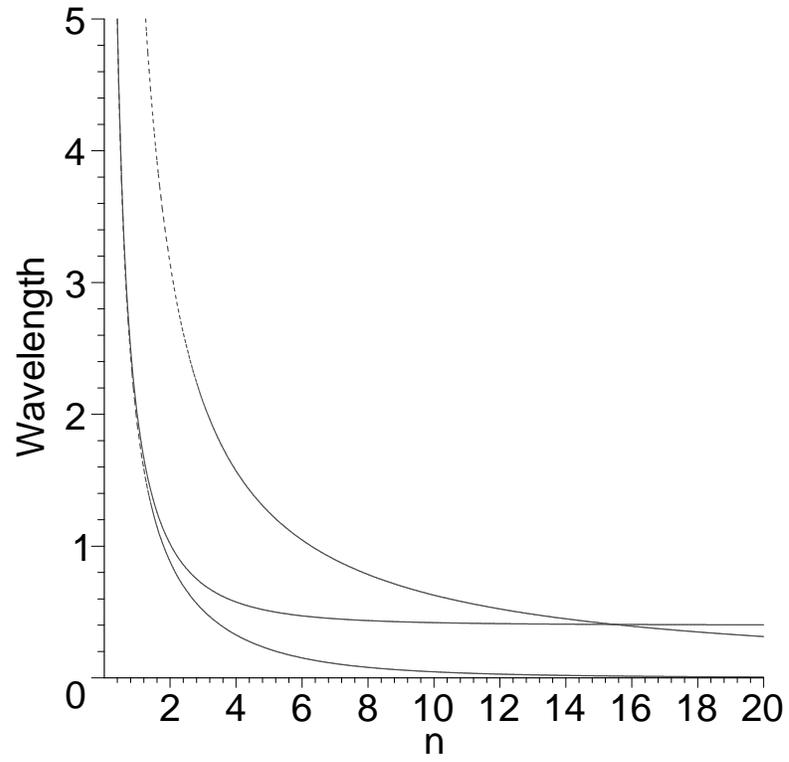}
\end{center}
\caption{Plot of Compton wavelengths associated
with the momentums $\kappa_3$ (solide), $p_3$ (dot) for $\beta$=0.01 and the usual one (dash-dot)
versus the quantum number $n$.}
\end{figure}

The potential vector in the presence of the plates is then given by%

\begin{align}
\mathbf{\hat{A}}_{a}(\mathbf{r},t)  &  =\sqrt{2\pi\hbar^{3}c\sqrt{\beta}}%
\frac{\hbar\pi}{a}\sum_{\underset{\gamma=\pm1}{n=-n_{\text{max}}}%
}^{n_{\text{max}}}\int\frac{d\mathbf{q}}{\left(  1+\beta\mathbf{p}^{2}\left(
a\right)  \right)  \sqrt{\arctan\sqrt{\beta}\mid\mathbf{p}\left(  a\right)
\mid}}\nonumber\\
&  \times\left\{  f_{\gamma}\left(  \mathbf{p}\left(  a\right)  ,\omega
\right)  \hat{a}_{\gamma}\left(  \mathbf{p}\left(  a\right)  \right)
+f^{\ast}\left(  \mathbf{p}\left(  a\right)  ,\omega\right)  \hat{a}_{\gamma
}^{\dagger}\left(  \mathbf{p}\left(  a\right)  \right)  \right\}  ,
\end{align}
where%

\begin{equation}
\mathbf{p}\left(  a\right)  =\mathbf{q}+\mathbf{p}_{3}\left(  n\right)  .
\end{equation}
The commutation relation between the creation and annihilation operators is
then affected by the solution $\left(  \ref{cond}\right)  .$ For our purpose
it suffice to we use the following approximation%

\begin{equation}
\left[  \hat{a}_{\gamma}\left(  \mathbf{p}\left(  a\right)  \right)  ,\hat
{a}_{\gamma^{\prime}}^{\dagger}\left(  \mathbf{p}^{\prime}\left(  a\right)
\right)  \right]  \simeq\frac{a}{\pi\hbar}\delta_{nn^{\prime}}\delta
_{\gamma\gamma^{\prime}}\delta(\mathbf{q}-\mathbf{q}^{\prime})+O\left(
\beta\right)  . \label{com}%
\end{equation}
The energy shift resulting from the presence of the plates is defined by the relation%

\begin{align}
\Delta E  &  =<0\mid\left(  \hat{H}(a)-\hat{H}\right)  \mid0>\nonumber\\
&  =\frac{1}{8\pi}\int d\mathbf{r}<0\mid\left\{  \left(  \partial
_{0}\mathbf{\hat{A}}_{a}\right)  ^{2}-\hat{\mathbf{A}}_{a}\Delta
\mathbf{\hat{A}}_{a}+\left(  \partial_{0}\mathbf{\hat{A}}\right)
^{2}-\mathbf{\hat{A}}\Delta\mathbf{\hat{A}}\right\}  \mid0>.
\end{align}
Performing the standard calculation we get%

\begin{equation}
\Delta E=\frac{cL^{2}}{8\pi\hbar^{2}\beta^{\frac{1}{2}}}\int d\mathbf{q}%
\left\{  \sum_{\underset{}{n=-n_{\text{max}}}}^{n_{\text{max}}}\frac
{\arctan\sqrt{\beta\left(  \mathbf{q}^{2}+p_{3}^{2}(n)\right)  }}%
{1+\beta\left(  \mathbf{q}^{2}+p_{3}^{2}(n)\right)  }-\int_{-\nu_{\text{max}}%
}^{\nu_{\text{max}}}d\nu\frac{\arctan\sqrt{\beta\left(  \mathbf{q}^{2}%
+p_{3}^{2}(\nu)\right)  }}{1+\beta\left(  \mathbf{q}^{2}+p_{3}^{2}%
(\nu)\right)  }\right\}  .
\end{equation}
From this expression it is easily seen that terms proportional to
$\beta^{n\geq1}$ in $p_{3}(n)$ and the omitted terms in the commutation
relation $\left(  \ref{com}\right)  $ will give negligible contributions
proportional to $\beta^{n\geq2}.$

Exchanging sums and integrals and defining the following quantity
\begin{equation}
G(\nu)=\frac{1}{\sqrt{\beta}}\int_{0}^{\infty}dx\frac{\arctan\sqrt
{\beta\left(  p_{3}^{2}\left(  \nu\right)  +x\right)  }}{1+\beta\left(
p_{3}^{2}\left(  \nu\right)  +x\right)  },
\end{equation}
the energy shift per unit area $\Delta\mathcal{E}=\frac{\Delta E}{L^{2}}$ is
given by%

\begin{equation}
\Delta\mathcal{E}=\frac{c}{4\pi\hbar^{2}}\left\{  \sum_{n=0}^{n_{\text{max}}%
}G(n)-\int_{0}^{\nu_{\max}}d\nu G(\nu)-\frac{1}{2}G(0)\right\}  .
\label{deltae}%
\end{equation}
With the aid of the variable $\rho=\frac{1}{\sqrt{\beta}}\arctan\sqrt
{\beta\left(  x+p_{3}^{2}\left(  \nu\right)  \right)  },$ the function
$G\left(  \nu\right)  $ is simply given by%

\begin{equation}
G(\nu)=\frac{2}{\sqrt{\beta}}\int_{\frac{1}{\sqrt{\beta}}\arctan p_{3}\left(
\nu\right)  \sqrt{\beta}}^{\frac{\pi}{2\sqrt{\beta}}}\tan\left(  \sqrt{\beta
}\rho\right)  \rho d\rho. \label{int}%
\end{equation}
Using the following expansion \cite{Guo}%

\begin{equation}
t\tan t=\sum\limits_{k=1}^{\infty}\frac{2^{2k}\left(  2^{2k}-1\right)  B_{k}%
}{\left(  2k\right)  !}t^{2k},\quad\left\vert t\right\vert <\frac{\pi}{2}%
\end{equation}
and performing the integral over $\rho$ we obtain%

\begin{equation}
G(\nu)=\sum\limits_{k=1}^{\infty}\frac{\beta^{k-1}2^{2k+1}\left(
2^{2k}-1\right)  B_{k}}{\left(  2k+1\right)  \left(  2k\right)  !}\left[
\left(  \frac{\pi}{2\sqrt{\beta}}\right)  ^{2k+1}-\left(  \frac{1}{\sqrt
{\beta}}\arctan p_{3}\left(  \nu\right)  \sqrt{\beta}\right)  ^{2k+1}\right]
\label{int02}%
\end{equation}
where $B_{k}$ are Bernoulli's numbers given by $B_{1}=\frac{1}{6},$
$B_{2}=\frac{1}{30},B_{3}=\frac{1}{42},\cdots\cite{grad}.$

It is important to note here that we have not introduced any cut-off as is the
case in the ordinary Casimir effect. The cut-off $\frac{1}{\sqrt{\beta}}$ is
implemented naturally in Eq.(\ref{int}). In Eq.$\left(  \ref{int02}\right)  $
the contributions for $n>$ $n_{\text{max }}$are negligeables compared to the
ones for $n\leq n_{\max}$ since $\frac{1}{\sqrt{\beta}}\arctan p_{3}\left(
\nu\right)  \sqrt{\beta}$ tends asymptotically to $\frac{\pi}{2\sqrt{\beta}}$
for $n>$ $n_{\text{max }}.$ For the rest of the calculation the first term is
irrelevant for our purpose and we ignore it. Then we can extend the summation
over $n$ and $\nu$ in Eq.$\left(  \ref{deltae}\right)  $ from $0$ to
$+\infty.$ Thus%

\begin{equation}
\Delta\mathcal{E}=\frac{c}{4\pi\hbar^{2}}\left\{  \sum_{n=0}^{\infty}%
G(n)-\int_{0}^{\infty}d\nu G(\nu)-\frac{1}{2}G(0)\right\}  . \label{deltaee}%
\end{equation}
At this stage we can use Euler formula \cite{grad}%

\begin{equation}
\int_{0}^{\infty}f(x)dx=\sum\limits_{n=0}^{\infty}f(n)-\frac{f(0)}{2}%
+\sum_{m=1}^{\infty}\frac{B_{2m}}{\left(  2m\right)  !}f^{\left(  2m-1\right)
}\left(  0\right)
\end{equation}
to obtain%

\begin{equation}
\Delta\mathcal{E}=-\frac{c}{4\pi\hbar^{2}}\sum_{m=1}^{\infty}\frac{B_{2m}%
}{\left(  2m\right)  !}G^{\left(  2m-1\right)  }\left(  0\right)  \label{true}%
\end{equation}
where $B_{2m}$ are Bernoulli numbers and $G^{\left(  l\right)  }\left(
0\right)  $ are derivatives of $G(\nu)$ at $\nu=0.$

Using the expression of $p_{3}\left(  \nu\right)  $ \ given by $\left(
\ref{cond}\right)  $ in $\left(  \ref{int02}\right)  $ we obtain to a first
order expansion in $\beta$ (Recall that the commutation relations are valid to
first order in $\beta$ )%

\begin{equation}
G(\nu)=-4B_{1}\left(  \frac{\hbar\pi\nu}{a}\right)  ^{3}+4\beta\left[
\frac{B_{1}}{3}+B_{2}\right]  \left(  \frac{\hbar\pi\nu}{a}\right)  ^{5}.
\end{equation}
Using $B_{1}=\frac{1}{6},$ $B_{2}=\frac{1}{30}$ we finally obtain%

\begin{equation}
G(\nu)=-\frac{2}{3}\left(  \frac{\hbar\nu\pi}{a}\right)  ^{3}+\frac{48}%
{135}\beta\left(  \frac{\hbar\nu\pi}{a}\right)  ^{5}.
\end{equation}
Then from Eq.(\ref{true}) we have %

\begin{equation}
\Delta\mathcal{E}=-\frac{c}{4\pi\hbar^{2}}\left[  \frac{B_{4}}{4!}G^{\left(
3\right)  }\left(  0\right)  +\frac{B_{6}}{6!}G^{\left(  5\right)  }\left(
0\right)  \right]  .
\end{equation}
Evaluating the derivatives at $\nu=0$ and using $B_{4}=\frac{1}{30},$
$B_{6}=\frac{691}{2730}$ we obtain%

\begin{equation}
\Delta\mathcal{E}=\hbar c\left\{  \frac{\pi^{2}}{720a^{3}}-\beta\frac
{691}{284275}\frac{\hbar^{2}\pi^{4}}{a^{5}}\right\}  .
\end{equation}
The force per unit surface $\mathcal{F}=\frac{\partial}{\partial a}%
\Delta\mathcal{E}$ generated by this energy is given by%

\begin{equation}
\mathcal{F}=-\hbar c\left\{  \frac{1}{240}\frac{\pi^{2}}{a^{4}}-\beta
\frac{691}{36855}\frac{\hbar^{2}\pi^{4}}{a^{6}}\right\}  . \label{force}%
\end{equation}
It is clear from this result, that for a fixed separation of the plates, the
Casimir force in the presence of a minimal length may be attractive of
repulsive depending on the value of the minimal length $\left(  \Delta
x\right)  _{\text{min}}=\hbar\sqrt{\beta}.$

The first term in Eq.$\left(  \ref{force}\right)  $ is the standard attractive
Casimir force \cite{cas} which, alone, is a source of instability. Indeed the
two plates systems can collapse to a one plate system.The second term which is
the correction arising from the presence of the minimal length is the
repulsive contribution to Casimir force and therefore provides the desired
stability of the two plates systems. This is important for the construction of
consistent Kaluza-Klein theories. The same results have been obtained by
\cite{nam} for the Casimir effect in $\kappa$-deformed theory and by
\cite{kemp4} for a particular implementation of the minimal length.

The condition for a quantum stability of the two plates systems gives the
following constraint%

\begin{equation}
\frac{\left(  \Delta x\right)  _{\text{min}}}{a}\sim0.15. \label{upperbound}%
\end{equation}
Using the experimentally accessible plates separations, which are of order
$100$ nm \cite{mohi}, we obtain%

\begin{equation}
\hbar\sqrt{\beta}\sim15\text{ nm.} \label{bound}%
\end{equation}
However for the force to remains attractive, as is usually observed, we have
the condition $\frac{\hbar\sqrt{\beta}}{a}\lesssim0.15$.

Figure 2 illustrate the variation of Casimir force for different values of the
minimal length. It is clear that this force becomes repulsive for
$\frac{\left(  \Delta x\right)  _{\text{min}}}{a}>0.15.$ Let us point that in
the plot $a$ is always greater than $\left(  \Delta x\right)  _{\text{min}}$
because the Casimir force for plates separation below the minimal length is
meaningful since the space below this scale is fuzzy and then experimentally inaccessible.

Before ending this section we note that the Casimir force in the presence of
one compactified eXtra dimension lies below the standard Casimir force
\cite{hossen}, while in the of a minimal length it lies above. Therefore we
can conclude that the effects of the minimal length and the eXtra dimensions
are opposites. This is expected from the beginning since the minimal length
squeeze the momentum space at high momentum and then the natural cut-off of
the model suppress the contributions of such momentum. Finally our treatment
along with the work in \cite{kemp4} contradict the one in \cite{hos} where the
Casimir force in the presence of a minimal length has been found to be a
discontinuous function of the plates separation, a result essentially due to
an appropriate geometric quantization between the plates.

\begin{figure}
\begin{center}
\includegraphics[height=10.0364cm, width=10.0364cm]{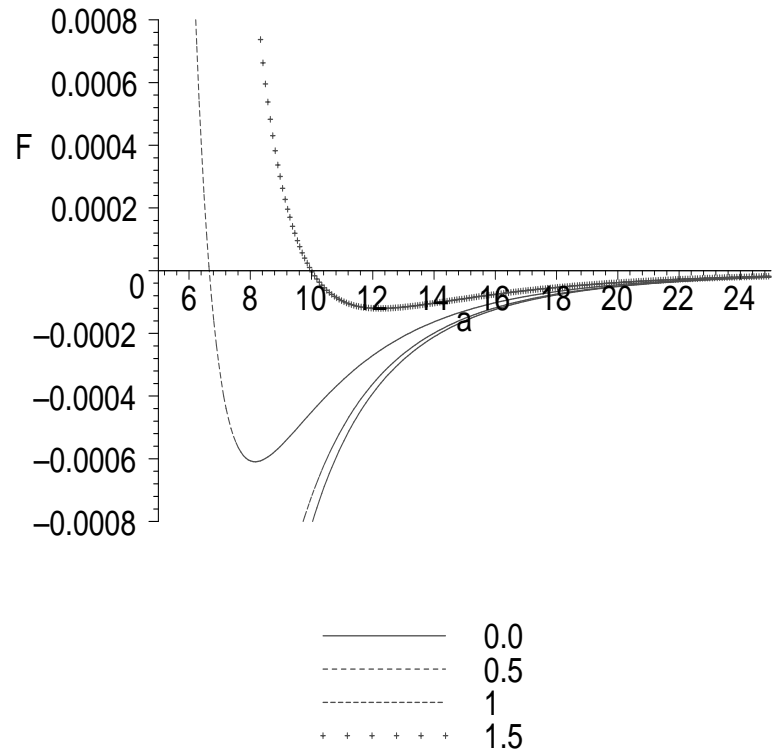}
\end{center}
\caption{Plot of Casimir Force F
[eV/nm$^3$] versus the plates separation $a$ 
[nm] for different values of the minimal length.}
\end{figure}

\section{Conclusion}

In this paper we considered the effect of minimal length on the Casimir force
between parallel plates. We shown that the minimal length acts like a natural
cut-off which suppress the contribution of unwanted high momentum. Using the
accessible plates separation used for an experimental calculation of Casimir
force we found an upper bound for the minimal length of the same order of the
size of one compactified eXtra dimension. However this bound is already
excluded from high precision mesurments and collider experiments \cite{hos01}
and then we recover the usual attractive character of Casimir force. The
Casimir force in the presence of minimal length in the context of a model with
one eXtra dimension is under investigation and will be published elsewhere.

\bigskip

\textbf{Knowledgment: The author thanks the referees for their pertinent and
valuable remarks.}

\end{document}